\documentclass[12pt]{article}
\usepackage{amsmath}
\usepackage{amssymb}
\usepackage{amsfonts}
\usepackage{latexsym}
\usepackage{color}
\usepackage{graphicx}

\catcode `\@=11 \@addtoreset{equation}{section}
\def\theequation{\arabic{section}.\arabic{equation}}
\catcode `\@=12

\newcommand{\be}{\begin{equation}}
\newcommand{\en}{\end{equation}}
\newcommand{\bea}{\begin{eqnarray}}
\newcommand{\ena}{\end{eqnarray}}
\newcommand{\beano}{\begin{eqnarray*}}
\newcommand{\enano}{\end{eqnarray*}}
\newcommand{\bee}{\begin{enumerate}}
\newcommand{\ene}{\end{enumerate}}

\newcommand{\N}{\mathfrak N}

\newcommand{\Hil}{{\cal H}}

\newcommand{\kt}{\rangle}
\newcommand{\br}{\langle}
\newcommand{\F}{{\cal F}}
\newcommand{\Lc}{{\cal L}}

\newcommand{\1}{1 \!\! 1}

\begin{document}

\thispagestyle{empty}

\vspace*{1cm}

\begin{center}
{\Large \bf Pseudo-fermions in an electronic loss-gain circuit}   \vspace{2cm}\\

{\large F. Bagarello}
\vspace{3mm}\\
  Dipartimento di Energia, Ingegneria dell'Informazione e Modelli Matematici,\\
Facolt\`a di Ingegneria, Universit\`a di Palermo, I-90128  Palermo, Italy\\
e-mail: fabio.bagarello@unipa.it
\vspace{2mm}\\
{\large G. Pantano}
\vspace{3mm}\\
  Facolt\`a di Ingegneria,
 Universit\`a di Palermo, I-90128  Palermo, Italy\\
e-mail: giuseppepantano.92@gmail.com
\end{center}

\vspace*{2cm}

\begin{abstract}
\noindent In some recent papers a loss-gain electronic circuit has been introduced and analyzed within the context of PT-quantum mechanics. In this paper we show that this circuit can be
analyzed using the formalism of the so-called pseudo-fermions. In particular we discuss the time behavior of the circuit, and we construct two biorthogonal
bases associated to the Liouville matrix $\Lc$ used in the treatment of the dynamics. We relate these bases to $\Lc$ and $\Lc^\dagger$, and we also show that a self-adjoint Liouville-like operator could be introduced in the game. Finally, we describe the time evolution of the circuit in an {\em
Heisenberg-like} representation, driven by a non self-adjoint hamiltonian.
\end{abstract}

\vspace{2cm}


\vfill

\newpage

\section{Introduction}

In some recent papers, \cite{bagpf1,bagpf2}, one of us (FB)  introduced the notion of {\em pseudo-fermions}, (PFs), arising from a deformed
version of the canonical anti-commutation relations (CAR). These PFs have been shown to be quite useful, mainly in connection with some specific quantum
mechanical systems, \cite{bagpf2}. Moreover, PFs are intrinsically related to a very nice functional structure, so that they appear also {\em mathematically appealing}.

Here we show how the same algebraic construction proposed for PFs can be useful also in the analysis of a completely different, classical,
system, i.e. an electronic circuit recently introduced in a series of recent papers, \cite{circu1,circu2,circu3}, in connection with PT-quantum
mechanics. In particular, by adopting our strategy, biorthogonal bases of the Hilbert space where the system lives, are generated, bases which
are therefore, somehow, {\em attached} to the circuit. Also, intertwining operators can be defined and two equivalent circuits, corresponding to the
adjoint version of the Liouvillian and to a third self-adjoint {\em similar} operators, can also be defined.

The paper is organized as follows: in the next section we introduce the electronic circuit and we derive the differential equations of motion.
We also list some results on PFs. In Section III we apply the pseudo-fermionic structure to the analysis of the dynamical behavior of the
circuit, adopting both the Schr\"odinger and an {\em Heisenberg-like} representation. We also consider other circuits which arise, because of
the existence of similarity transformations, starting from the original Liouvillian. Section IV contains our conclusions, while a different approach to
the dynamical behavior of the circuit is sketched in the Appendix.

\section{Stating the problem and first considerations}

\begin{figure}[h]
\begin{center}
\includegraphics[height=5.2cm, width=9.5cm]{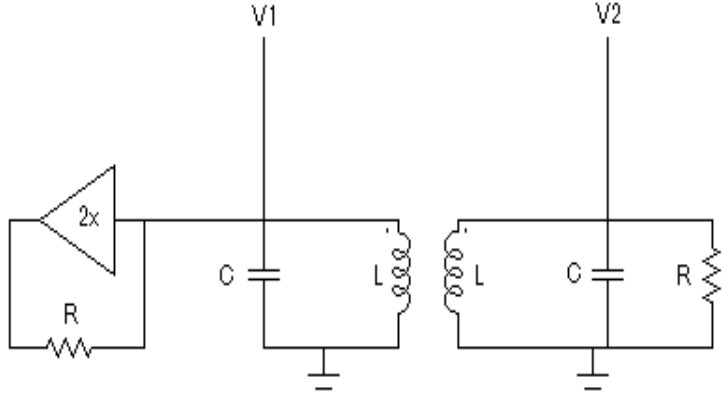}\hfill\\
\caption{\label{circuito}\footnotesize The two-components circuit.}
\end{center}
\end{figure}

In \cite{circu1,circu2,circu3} the authors, with the aim of discussing a suitable interplay between loss and gain in a two-components circuit,
 introduced a very simple model, see Figure \ref{circuito}, consisting in two different parts, interacting via a mutual inductance. The physical interest of this circuit is that it produces a concrete system which, apparently, seems to produce an arbitrary fast dynamics. The reason for that is that the time evolution is not unitarily implemented, while it is tuned by a suitably chosen non-hermitian hamiltonian.

Calling $V_j(t)$ and $I_j(t)$, $j=1,2$, the potential and the current for the $j$-th component  of the circuit, the following equations are
easily deduced: \be\left\{
\begin{array}{ll}
V_1(t)= L\dot I_1(t)+M\dot I_2(t),\\ V_2(t)= L\dot I_2(t)+M\dot I_1(t),\\
I_1(t)=\frac{1}{R}\,V_1(t)-C\dot V_1(t), \\ I_2(t)=-\,\frac{1}{R}\,V_2(t)-C\dot V_2(t).
\label{21}\end{array}
\right.
\en
If we now introduce $\omega_0=\frac{1}{\sqrt{LC}}$, $\tau=\omega_0 t$, $\mu=\frac{M}{L}$, $\gamma=\frac{1}{R}\sqrt{\frac{L}{C}}$ and $\alpha=\frac{1}{1-\mu^2}$, where we have assumed that $\mu\neq\pm 1$, the following equation are deduced for $V_1(\tau)$ and $V_2(\tau)$:
\be\left\{
\begin{array}{ll}
V_1''(\tau)= -\alpha V_1(\tau)+\alpha\mu V_2(\tau)+\gamma V_1'(\tau),\\ V_2''(\tau)= -\alpha V_2(\tau)+\alpha\mu V_1(\tau)-\gamma V_2'(\tau).
\label{22}\end{array}
\right.
\en
Here the {\em prime} is the derivative with respect to $\tau$, which is clearly proportional to the ordinary time derivative. We will see that these equations can be rewritten as two uncoupled, fourth-order, differential equations in the Appendix. Here we are more interested in considering them from a different point of view. For that, we introduce the vector $\Psi(\tau)$ and the matrix $\Lc$ as follows:
\be
\Lc=\left(
      \begin{array}{cccc}
        0 & 0 & 1 & 0 \\
        0 & 0 & 0 & 1 \\
        -\alpha & \alpha\mu & \gamma & 0 \\
        \alpha\mu & -\alpha & 0 & -\gamma \\
      \end{array}
    \right),\qquad \Psi(\tau)=\left(
                                \begin{array}{c}
                                  V_1(\tau) \\
                                  V_2(\tau) \\
                                  V_1'(\tau) \\
                                  V_2'(\tau) \\
                                \end{array}
                              \right).
    \label{23}\en
Then it is clear that (\ref{22}) can be rewritten as \be \Psi'(\tau)=\Lc\Psi(\tau), \label{24}\en which could be still written as
$i\Psi'=H_{eff}\Psi$, simply by introducing a $4\times4$ matrix $H_{eff}=i\Lc$, \cite{circu2}. This  can be seen as a Schr\"odinger-like equation, with
$H_{eff}$ manifestly not self-adjoint. However, it should be stressed that this is not really so simple, since the four components of the
vector $\Psi(\tau)$, contrarily to what happens in a general quantum mechanical system, are related among them: the third component, $V_1'(\tau)$, is infact the $\tau$-derivative of the
first one. It might be interesting to
notice that going from (\ref{22}) to (\ref{24}) is nothing but doubling the number of variables to rewrite a second order differential equation
as a set of two first-order differential equations, which is a standard procedure in the mathematical literature.

The analysis of the circuit in Figure \ref{circuito} was used in \cite{circu2} as a prototype model which bypass, as the authors suggest, the lower bound imposed by the bandwidth theorem. This is not our main interest here: in fact, we are more interested in showing that PFs can be useful in the general treatment of equation (\ref{24}), treatment which will naturally produce, as we will show, more {\em equivalent} circuits.

 Before beginning our analysis, we need to recall few useful
and interesting facts on PFs.

\subsection{The pseudo-fermionic structure}

We limit our analysis of PFs to one and two dimensions. The extension to higher dimensions is straightforward, and it will not be given here,
since will not be useful for us. We begin with $d=1$. The starting point is a modification of the CAR $\{c,c^\dagger\}=c\,c^\dagger+c^\dagger\,c=\1$,
$\{c,c\}=\{c^\dagger,c^\dagger\}=0$, between two operators, $c$ and $c^\dagger$, acting on a two-dimensional Hilbert space $\Hil$. The CAR are
replaced here by the following rules: \be \{a,b\}=\1, \quad \{a,a\}=0,\quad \{b,b\}=0, \label{220}\en where the interesting situation is when
$b\neq a^\dagger$. These rules automatically imply that a non zero vector, $\varphi_0$, exists in $\Hil$ such that $a\,\varphi_0=0$, and that a
second non zero vector, $\Psi_0$, also exists in $\Hil$ such that $b^\dagger\,\Psi_0=0$, \cite{bagpf1}.

Let us now introduce the following non zero vectors \be \varphi_1:=b\varphi_0,\quad \Psi_1=a^\dagger \Psi_0, \label{221}\en as well as the non
self-adjoint operators \be N=ba,\quad \N:=N^\dagger=a^\dagger b^\dagger. \label{222}\en We further introduce the self-adjoint operators
$S_\varphi$ and $S_\Psi$ via their action on a generic $f\in\Hil$: \be S_\varphi f=\sum_{n=0}^1\br\varphi_n,f\kt\,\varphi_n, \quad S_\Psi
f=\sum_{n=0}^1\br\Psi_n,f\kt\,\Psi_n. \label{223}\en Hence we get the following results,   whose proofs are straightforward:

\begin{enumerate}

\item \be a\varphi_1=\varphi_0,\quad b^\dagger\Psi_1=\Psi_0.\label{224}\en

\item \be N\varphi_n=n\varphi_n,\quad \N\Psi_n=n\Psi_n,\label{225}\en
for $n=0,1$.

\item If the normalizations of $\varphi_0$ and $\Psi_0$ are chosen in such a way that $\left<\varphi_0,\Psi_0\right>=1$,
then \be \left<\varphi_k,\Psi_n\right>=\delta_{k,n},\label{226}\en for $k,n=0,1$.

\item $S_\varphi$ and $S_\Psi$ are bounded, strictly positive, self-adjoint, and invertible. They satisfy
\be \|S_\varphi\|\leq\|\varphi_0\|^2+\|\varphi_1\|^2, \quad \|S_\Psi\|\leq\|\Psi_0\|^2+\|\Psi_1\|^2,\label{227}\en \be S_\varphi
\Psi_n=\varphi_n,\qquad S_\Psi \varphi_n=\Psi_n,\label{228}\en for $n=0,1$, as well as $S_\varphi=S_\Psi^{-1}$. Moreover, the following
intertwining relations \be S_\Psi N=\N S_\Psi,\qquad S_\varphi \N=N S_\varphi,\label{229}\en hold.

\end{enumerate}

The above formulas show that (i) $N$ and $\N$ behave (almost) like fermionic number operators, having eigenvalues 0 and 1; (ii) their related eigenvectors
are respectively the vectors of $\F_\varphi=\{\varphi_0,\varphi_1\}$ and $\F_\Psi=\{\Psi_0,\Psi_1\}$; (iii) $a$ and $b^\dagger$ are lowering
operators for $\F_\varphi$ and $\F_\Psi$ respectively; (iv) $b$ and $a^\dagger$ are rising operators for $\F_\varphi$ and $\F_\Psi$
respectively; (v) the two sets $\F_\varphi$ and $\F_\Psi$ are biorthonormal; (vi) the {\em very well-behaved} operators $S_\varphi$ and
$S_\Psi$ maps $\F_\varphi$ in $\F_\Psi$ and viceversa; (vii) $S_\varphi$ and $S_\Psi$ intertwine between operators which are not self-adjoint. Another interesting feature is the following:
since the square roots of $S_\Psi$ and $S_\varphi$ surely exist, from the first equation in (\ref{229}) we get
$$
S_\Psi^{1/2}NS_\Psi^{-1/2}=S_\Psi^{-1/2}\N S_\Psi^{1/2}=\left(S_\Psi^{1/2}NS_\Psi^{-1/2}\right)^\dagger,
$$
which states that $\hat n:=S_\Psi^{1/2}NS_\Psi^{-1/2}$ is a self-adjoint operator, similar to $N$ (and to $\N$, since $\hat n=S_\varphi^{1/2}\N S_\varphi^{-1/2}$).

\subsubsection{A two-dimensional extension}\label{sectpf2d}

Let $(a_j,b_j)$ be two pairs of pseudo-fermionic operators, $\{a_j,b_j\}=\1$, $a_j^2=b_j^2=0$, $j=1,2$, satisfying also the following independence relation: $\{a_j^\sharp,b_k^\sharp\}=0$, for $j\neq k$, and $x^\sharp=x$ or $x^\sharp=x^\dagger$. Let $\varphi_{0,0}$ be a vector annihilated by $a_1$ and $a_2$: $a_1\varphi_{0,0}=a_2\varphi_{0,0}=0$. Then, introducing $\varphi_{1,0}=b_1\varphi_{0,0}$, $\varphi_{0,1}=b_2\varphi_{0,0}$, and $\varphi_{1,1}=b_1\,b_2\varphi_{0,0}$, the set $\F_\varphi=\{\varphi_{k,l}, \,k,l=0,1\}$ is a basis for $\Hil={\Bbb C}^4$ of eigenstates of $N_1=b_1a_1$ and $N_2=b_2a_2$: $N_1\varphi_{k,l}=k\varphi_{k,l}$, and $N_2\varphi_{k,l}=l\varphi_{k,l}$. Similar results as those deduced in the one-dimensional case can be recovered also here. For instance, a biorthogonal basis of $\Hil$, $\F_\Psi$, can be found, and these new vectors are eigenstates of $N_j^\dagger$, $j=1,2$. Also, intertwining operators mapping $\F_\Psi$ into $\F_\varphi$ and viceversa can again be defined.

\vspace{2mm}

We refer to \cite{bagpf1} for further remarks and consequences of these definitions. In particular, for instance, it is shown that $\F_\varphi$
and $\F_\Psi$ are automatically Riesz bases for $\Hil$, and the relations between fermions and PFs are discussed.

\section{Pseudo-fermions from the circuit}

In this section we will work under the following useful requirements: \be \rho:=\gamma^4+4\alpha^2\mu^2-4\alpha\gamma^2>0,\quad
\gamma^2-2\alpha>0,\quad 0\leq\mu^2<1. \label{31}\en These conditions allow us to check that the eigenvalues of $\Lc$ are all different and
reals. In particular, calling $l_1=-\frac{1}{\sqrt{2}}(\gamma^2-2\alpha-\sqrt{\rho})^{1/2}$,
$l_2=\frac{1}{\sqrt{2}}(\gamma^2-2\alpha-\sqrt{\rho})^{1/2}=-l_1$, $l_3=-\frac{1}{\sqrt{2}}(\gamma^2-2\alpha+\sqrt{\rho})^{1/2}$ and
$l_4=\frac{1}{\sqrt{2}}(\gamma^2-2\alpha+\sqrt{\rho})^{1/2}=-l_3$, we deduce that $l_3 <  l_1 <0 < l_2 < l_4$. Then, if we introduce
$\tilde\Lc=\Lc-l_3\1$, its eigenvalues $\lambda_j$, $j=0,1,2,3$, are easily found: $\lambda_0=0$, $\lambda_1=l_1-l_3$, $\lambda_2=l_2-l_3$,
$\lambda_3=l_4-l_3$,  and the following hold:
$$
0=\lambda_0<\lambda_1<\lambda_2<\lambda_3=\lambda_1+\lambda_2.
$$

Let us introduce the matrices
$$
A_1=\left(
      \begin{array}{cccc}
        0 & 1 & 0 & 0 \\
        0 & 0 & 0 & 0 \\
        0 & 0 & 0 & 1 \\
        0 & 0 & 0 & 0 \\
      \end{array}
    \right), \qquad A_2=\left(
      \begin{array}{cccc}
        0 & 0 & 1 & 0 \\
        0 & 0 & 0 & -1 \\
        0 & 0 & 0 & 0 \\
        0 & 0 & 0 & 0 \\
      \end{array}
    \right).
$$
They satisfy the following CAR: $A_j^2=0$, and $\{A_j,A_k^\dagger\}=\delta_{j,k}\1$, $j,k=1,2$. We further introduce the following self-adjoint
operator: $H_0=\lambda_1\,A_1^\dagger A_1+\lambda_2\,A_2^\dagger A_2$, whose eigenstates are
$$
\Phi_{0,0}=\left(
             \begin{array}{c}
               1 \\
               0 \\
               0 \\
               0 \\
             \end{array}
           \right), \quad \Phi_{1,0}=\left(
             \begin{array}{c}
               0 \\
               1 \\
               0 \\
               0 \\
             \end{array}
           \right), \quad \Phi_{0,1}=\left(
             \begin{array}{c}
               0 \\
               0 \\
               1 \\
               0 \\
             \end{array}
           \right), \quad \Phi_{1,1}=\left(
             \begin{array}{c}
               0 \\
               0 \\
               0 \\
               1 \\
             \end{array}
           \right).
$$
They are orthonormal and satisfy the eigenvalue equation $H_0\Phi_{k,n}=(k\lambda_1+n\lambda_2)\Phi_{k,n}$, $k,n=0,1$. Moreover,  $\Phi_{1,0}=A_1^\dagger\Phi_{0,0}$, $\Phi_{0,1}=A_2^\dagger\Phi_{0,0}$, and $\Phi_{1,1}= A_1^\dagger A_2^\dagger\Phi_{0,0}$.

It is possible to show that $H_0$ and $\tilde\Lc$ are related by an intertwining operator $T$. In fact we can deduce
\be
\tilde\Lc\, T=T\, H_0,
\label{32}\en
where $T$ is the following matrix:
$$
T=\left(
    \begin{array}{cccc}
      \delta_{21}t_{21} & \delta_{22}t_{22} & \delta_{23}t_{23} & \delta_{24}t_{24} \\
      t_{21} & t_{22} & t_{23} & t_{24} \\
      l_3\delta_{21}t_{21} &  l_1 \delta_{22}t_{22} &  l_2 \delta_{23}t_{23} & l_4\delta_{24}t_{24} \\
      l_3t_{21} &  l_1 t_{22} &  l_2 t_{23} & l_4t_{24} \\
    \end{array}
  \right).
$$
Consequences of (\ref{32}) will be considered below. Here the following quantities have been introduced:
$$
\delta_{21}=\frac{1}{2\alpha\mu}(\gamma^2+\sqrt{\rho}-2\gamma l_4),\quad \delta_{22}=\frac{1}{2\alpha\mu}(\gamma^2-\sqrt{\rho}-2\gamma l_2 ),
$$
$$
\delta_{23}=\frac{1}{2\alpha\mu}(\gamma^2-\sqrt{\rho}+2\gamma l_2 ),\quad \delta_{24}=\frac{1}{2\alpha\mu}(\gamma^2+\sqrt{\rho}+2\gamma l_4).
$$
Since
$$
\det(T)=-\,\frac{4\rho\,l_4\, l_2 }{\alpha^2\mu^2}\,t_{21}t_{22}t_{23}t_{24},
$$
it is clear that $\det(T)$ is always non zero if the four $t_{2,j}$, $j=1,2,3,4$, are non zero. In this case, $T$ is invertible and the
previous intertwining relation becomes $\tilde\Lc=T\,H_0\,T^{-1}$: as a consequence, the non self-adjoint  Liouvillian
$\Lc=\tilde\Lc+l_3\1=T\,(H_0+l_3\1)\,T^{-1}$ associated to the circuit in Figure \ref{circuito} is similar to the self-adjoint adjoint
hamiltonian $H_0$ (plus $l_3\1$), whose eigenvalues and eigenvectors are given above.

\subsection{Consequences of the pseudo-fermionic settings}

What discussed in Section II suggests to introduce now the  operators
$a_j=TA_jT^{-1}$ and $b_j=TA_j^\dagger T^{-1}$, $j=1,2$, since in this way $\Lc$ can be written as $\Lc=\lambda_1N_1+\lambda_2N_2+l_3\1$,
where, as in Section \ref{sectpf2d}, we have introduced $N_j=b_ja_j$. It is obvious that $(a_j,b_j)$ are pseudo-fermionic operators:
$a_j^2=b_j^2=0$, $\{a_j,b_k\}=\1\delta_{j,k}$, $j,k=0,1$. The eigenstates of $\Lc$ can be constructed from the vacuum of $a_j$, $\varphi_{0,0}$
satisfying $a_j\varphi_{0,0}=0$, $j=1,2$: $\varphi_{1,0}=b_1\varphi_{0,0}$, $\varphi_{0,1}=b_2\varphi_{0,0}$,
$\varphi_{1,1}=b_1b_2\varphi_{0,0}$. Then
\be
\Lc\varphi_{k,n}=(k\lambda_1+n\lambda_2+l_3)\varphi_{k,n},
\label{33}\en
 $k,n=0,1$. It is now easy to check that there exists a relation between the vectors $\varphi_{k,n}$ and $\Phi_{k,n}$. In fact we have  $\varphi_{k,n}=T\Phi_{k,n}$, $k,n=0,1$. Needless to say, the set $\F_\varphi=\{\varphi_{k,n}\}$
 is a basis for $\Hil$. However, since $T$ is not unitary, $\F_\varphi$ is not an o.n. basis. It is very easy now to find a second
 set of vectors,  $\F_\Psi=\{\Psi_{k,n},\,k,n=0,1\}$, which is a new basis, biorthogonal to $\F_\varphi$. For that it is sufficient to
 introduce the vectors like this: $\Psi_{k,n}=(T^{-1})^\dagger\Phi_{k,n}$, $k,n=0,1$, which surely exist in our hypotheses, since $T$ is invertible.
 We can check the following facts:
 \begin{enumerate}

 \item As already stated, $\F_\varphi$ and $\F_\Psi$ are biorthogonal: $\langle\Psi_{k,n},\varphi_{l,m}\rangle=\delta_{k,l}\delta_{n,m}$.

 \item $\F_\varphi$ and $\F_\Psi$ satisfy the following resolutions of the identity: $\sum_{k,n}|\Psi_{k,n}\rangle\langle\varphi_{k,n}|=\1$ and  $\sum_{k,n}|\varphi_{k,n}\rangle\langle\Psi_{k,n}|=\1$.

 \item Defining an operator $S_\varphi$ as $S_\varphi f=\sum_{k,n}\langle\varphi_{k,n},f\rangle \varphi_{k,n}$, this can be written as $S_\varphi=T\,T^\dagger$. Hence it is strictly positive and, clearly, self-adjoint.

\item Analogously, defining an operator $S_\Psi$ as $S_\Psi f=\sum_{k,n}\langle\Psi_{k,n},f\rangle \Psi_{k,n}$,
 it turns out that $S_\Psi=S_\varphi^{-1}=(T^\dagger)^{-1}T^{-1}$.

\item The vectors $\Psi_{k,n}$ are eigenstates of $\tilde\Lc^\dagger$ and, consequently, of $\Lc^\dagger$:
\be
 \Lc^\dagger\Psi_{k,n}=(k\lambda_1+n\lambda_2+l_3)\Psi_{k,n}, \label{34}\en $k,n=0,1$. Hence $\Lc$ and $\Lc^\dagger$ are isospectral, as
expected. This is, in fact, a simple consequence of the fact that these two operators are related by an intertwining operator, $T$, as we
will see in Section \ref{sectIII2}.


\end{enumerate}

Let us now go back to equation (\ref{24}), $\Psi'(\tau)=\Lc\Psi(\tau)$. We look for a solution of this equation as the following linear combination of vectors of $\F_\varphi$: $\Psi(\tau)=c_{0,0}(\tau)\varphi_{0,0}+c_{1,0}(\tau)\varphi_{1,0}+c_{0,1}(\tau)\varphi_{0,1}+c_{1,1}(\tau)\varphi_{1,1}$. This is a natural choice, since $\varphi_{k,n}$ are eigenstates of $\Lc$. The analytical expressions of the various $c_{i,j}(\tau)$ can be easily deduced by inserting the expansion above for $\Psi(\tau)$ in (\ref{24}), and using the biorthogonality of $\F_\varphi$ and $\F_\Psi$. $\Psi(\tau)$ is found to be
\be
\Psi(\tau)=e^{l_3\tau}\left(c_{0,0}(0)\varphi_{0,0}+e^{\lambda_1\tau}c_{1,0}(0)\varphi_{1,0}+e^{\lambda_2\tau}c_{0,1}(0)\varphi_{0,1}+
e^{(\lambda_1+\lambda_2)\tau}c_{1,1}(0)\varphi_{1,1}\right),
\label{35}\en
where the different $c_{i,j}(0)$ are fixed by the initial conditions. We adopt here the choice in \cite{circu2}: $V_1(0)=V_2(0)=I_2(0)=0$, and $I_1(0)=i_1$. Using (\ref{21}) we find that $\Psi(0)^T=\left(
                                     \begin{array}{cccc}
                                       0 & 0 & -i_1/C & 0 \\
                                     \end{array}
                                   \right)
$, so that $$ \left\{
\begin{array}{ll}
c_{0,0}=-\frac{-l_4(\delta_{22}-\delta_{23})+ l_2 (\delta_{22}+\delta_{23}-2\delta_{24})i_1}{\sigma\, t_{21}},\quad
c_{1,0}=-\frac{- l_2 (\delta_{21}-\delta_{24})+l_4(\delta_{21}+\delta_{24}-2\delta_{23})i_1}{\sigma \,t_{22}},\\
c_{0,1}=\frac{ l_2 (\delta_{21}-\delta_{24})+l_4(\delta_{21}+\delta_{24}-2\delta_{22})i_1}{\sigma \,t_{23}},\qquad \,\,
c_{1,1}=\frac{l_4(\delta_{22}-\delta_{23})+ l_2 (\delta_{22}+\delta_{23}-2\delta_{21})i_1}{\sigma \,t_{24}}  ,
\end{array}
\right.$$ where we have introduced $$\sigma:=C[(l_4^2+ l_2 ^2)(\delta_{22}-\delta_{23})\delta_{21}\delta_{24}-2l_4
 l_2 (-2(\delta_{22}\delta_{23}+\delta_{21}(\delta_{22}+\delta_{23}-2\delta_{24})+\delta_{22}\delta_{24}+\delta_{23}\delta_{24})]. $$ To
simplify the notation we have written $c_{k,l}$ instead of $c_{k,l}(0)$. It is now not difficult, using (\ref{23}),  to deduce the expression
for $V_j(\tau)$ and for $I_j(\tau)$:
$$\left\{
\begin{array}{ll}
 V_1(\tau)=e^{l_3\tau}\left(c_{0,0}\delta_{21}t_{21}+e^{\lambda_1\tau}c_{1,0}\delta_{22}t_{22}+e^{\lambda_2\tau}c_{0,1}\delta_{23}t_{23}+
e^{(\lambda_1+\lambda_2)\tau}c_{1,1}\delta_{24}t_{24}\right),\\
V_2(\tau)=e^{l_3\tau}\left(c_{0,0}t_{21}+e^{\lambda_1\tau}c_{1,0}t_{22}+e^{\lambda_2\tau}c_{0,1}t_{23}+
e^{(\lambda_1+\lambda_2)\tau}c_{1,1}t_{24}\right),\\
I_1(\tau)= e^{l_3\tau}( c_{0,0}\delta_{21}t_{21}(\frac1R+Cl_4)+e^{\lambda_1\tau}c_{1,0}\delta_{22}t_{22}(\frac1R+C l_2 )
+\\\qquad\quad+ \,e^{\lambda_2\tau}c_{0,1}\delta_{23}t_{23}(\frac1R-C l_2 ) +e^{(\lambda_1+\lambda_2)\tau}c_{1,1}\delta_{24}t_{24}
(\frac1R-Cl_4)),\\
I_2(\tau)= e^{l_3\tau}( c_{0,0}t_{21}(-\frac1R+Cl_4)+e^{\lambda_1\tau}c_{1,0}t_{22}(-\frac1R+C l_2 ) +     \\
\qquad \quad +\, e^{\lambda_2\tau}c_{0,1}t_{23}(-\frac1R-C l_2 ) +e^{(\lambda_1+\lambda_2)\tau}c_{1,1}t_{24}(-\frac1R-Cl_4) ).
\end{array}
\right. $$

\vspace{2mm}

Let us introduce now the power of the two sub-circuits as $P_j(\tau):=V_j(\tau)I_j(\tau)$, $j=1,2$. Because of (\ref{21}) we can write
$$
P_j(\tau)=\frac{(-1)^{j+1}}{R}\,V_j^2(\tau)-CV_j'(\tau)V_j(\tau),
$$
$j=1,2$. The asymptotic behavior of $P_j(\tau)$ can be deduced from the expressions above for $V_j(\tau)$, and we can check that it only
depends on the sum $\lambda_1+\lambda_2+l_3=-l_3$, which is always positive. Indeed we have, for very large $\tau$, $$ \left\{\begin{array}{ll}
    P_1(\tau)\simeq
e^{2(\lambda_1+\lambda_2+l_3)\tau}\left(c_{1,1}(0)t_{24}\delta_{24}\right) ^2\left( \frac{1}{R}-Cl_4\right)\\
    P_2(\tau)\simeq
e^{2(\lambda_1+\lambda_2+l_3)\tau}\left(c_{1,1}(0)t_{24}\right) ^2\left( -\frac{1}{R}-Cl_4\right )  \label{30}\end{array} \right. $$ Both these
functions, therefore, diverge. However, if $\frac{1}{R}-Cl_4>0$ and $-\frac{1}{R}-Cl_4<0$, $P_1(\tau)$ diverges to $+\infty$, while $P_2(\tau)$
diverges to $-\infty$. This different behavior could be seen as an evidence of a gain (for the first sub-circuit) and a loss (for the second
sub-circuit), see Figure \ref{circuito}. It is interesting to observe that the two conditions can be written as \be
-\,\frac{1}{RC}<l_4<\frac{1}{RC}, \label{51}\en which has an interesting interpretation: in order for the power of the two coupled sub-circuits
to describe loss and gain, $l_4$ must be between the two damping constants of the two sub-circuits.

\vspace{2mm}

A similar analysis can be carried out if we consider the energy of the two sub-circuits, as in \cite{circu2}:
$E_n(\tau)=\frac12CV_n(\tau)^2+\frac12LI_n(\tau)^2$, $n=1,2$. Using equations (\ref{21}), putting $\omega_0=\frac{1}{\sqrt{LC}}$ and
$\omega_p=\frac{1}{RC}$, we can write $E_1(\tau)=\frac12 LC^2(V_1(\tau)^2(\omega_0^2+\omega_p^2)-V_1'(\tau)^2)$ and $E_2(\tau)=\frac12
LC^2(V_2(\tau)^2(\omega_0^2-\omega_p^2)-V_2'(\tau)^2)$. It is now possible, in principle, analyze $E_n(\tau)$ for all $\tau$. However, here, we
will limit ourselves to consider the asymptotic behavior for $\tau$ very large. Repeating the same steps as above, we deduce that $E_1(\tau)$
diverges to $+\infty$ if $\omega_0^2+\omega_p^2-l_4^2>0$, while $E_2(\tau)$ diverges to $-\infty$ if $\omega_0^2-\omega_p^2-l_4^2<0$. They are
both satisfied if $\sqrt{\omega_0^2-\omega_p^2}<l_4<\sqrt{\omega_0^2+\omega_p^2}$, which is very similar to (\ref{51}). The only difference is
in the appearance of both $\omega_0$ and $\omega_p$, which therefore both play a role in this analysis: the eigenvalue $l_4$ must belong to a
suitable neighborhood of $\omega_0$, with a width fixed by $\omega_p$.

\subsection{On $\Lc^\dagger$}\label{sectIII2}

We have seen that, adopting our {\em pseudo-fermionic strategy}, a second natural operator, other that $\Lc$, appears in the game. This
operator, $\Lc^\dagger$, can be directly related to $\Lc$ simply recalling that $\Lc=\tilde\Lc+l_3\1$ and that $\tilde\Lc=TH_0T^{-1}$. In fact,
these simple equalities imply the following \be \Lc=T\left(H_0+l_3\1\right)T^{-1},\quad\mbox{ so that }\quad
\Lc^\dagger={T^{-1}}^\dagger\left(H_0+l_3\1\right){T}^\dagger. \label{extra1}\en Therefore, recalling that $S_\varphi=TT^\dagger$, we conclude
that $\Lc=S_\varphi\Lc^\dagger S_\varphi^{-1}$ or, equivalently, that $\Lc S_\varphi=S_\varphi\Lc^\dagger$. This last equation is a typical
intertwining relation, \cite{intop}, relating $\Lc$ and $\Lc^\dagger$ by means of the intertwining operator $S_\varphi$. Among the other
consequences of this relation, a crucial one is that the eigenvalues of $\Lc$ and $\Lc^\dagger$ should coincide, as it actually happens in our
concrete model. Moreover, the related eigenvectors of $\Lc$ and $\Lc^\dagger$ should be somehow related by $S_\varphi$. Again, this is exactly
what happens here. In fact, recalling that $\varphi_{k,n}=T\Phi_{k,n}$ and that $\Psi_{k,n}=(T^{-1})^\dagger\Phi_{k,n}$, $k,n=0,1$, we deduce
that $\varphi_{k,n}=S_\varphi\Psi_{k,n}$, $k,n=0,1$, as expected. It could be worth stressing that these results are not peculiar of the model
we are considering here; they appear everywhere when pseudo-fermions (or pseudo-bosons, \cite{bagpbnew}), are involved.

Going back to $\Lc=S_\varphi\Lc^\dagger S_\varphi^{-1}$, this means that, \cite{bagpbnew}, $\Lc$ is crypto-hermitian with respect to
$S_\varphi^{-1}$. This fact has a lot of consequences, which are described in \cite{bagpbnew}. We should probably stress that all the
mathematical difficulties which we are forced to consider in \cite{bagpbnew}, here do not appear, since we are working with intrinsically
bounded operators (finite-dimensional matrices!).

We can now replace $\Lc=S_\varphi\Lc^\dagger S_\varphi^{-1}$ in the differential equation (\ref{24}). Defining further a new vector
$\eta(\tau):=S_\varphi^{-1}\Psi(\tau)$, we get \be \eta'(\tau)=\Lc^\dagger\eta(\tau),\label{37}\en which can be seen as the differential
equation {\em generated by}  $\Lc^\dagger$, whose solution can be easily found, $\eta(\tau)=S_\varphi^{-1}\Psi(\tau)$, once the solution of
(\ref{24}) is known. Of course, we could reverse the conclusion: suppose we have solved (\ref{37}). Then, the solution $\Psi(\tau)$ of (\ref{24}) is deduced by $\Psi(\tau)=S_\varphi\eta(\tau)$.

\vspace{3mm}

The above procedure does not clarify the {\em electronic} meaning of $\Lc^\dagger$. Then, it is interesting to set up a different
procedure. For this reason, we assume that the four dimensional vector $X(\tau)$, with $X^T(\tau)=(x_1(\tau),x_2(\tau),x_3(\tau),x_4(\tau))$,
satisfies the differential equation $ X'(\tau)=\Lc^\dagger X(\tau)$. After some minor manipulations, and recalling that
$\alpha=\frac{1}{1-\mu^2}$, we get the following set of equations for $x_j(\tau)$: \be\left\{
\begin{array}{ll}
x_3(\tau)= - x_1'(\tau)-\mu  x_2'(\tau),\\
x_4(\tau)= - x_2'(\tau)-\mu  x_1'(\tau),\\
x_1(\tau)=-\gamma x_3(\tau)+ x_3'(\tau)\\
x_2(\tau)=\gamma x_4(\tau)+ x_4'(\tau). \label{311}\end{array} \right. \en This set of equations are analytically very close to that in
(\ref{21}). In particular, they even coincide if we make the following identifications: $x_1(\tau)\leftrightarrow I_1(\tau)$,
$x_2(\tau)\leftrightarrow I_2(\tau)$, $x_3(\tau)\leftrightarrow -V_1(\tau)$ and $x_4(\tau)\leftrightarrow -V_2(\tau)$. The only price we have
to pay is that we also need to fix $L=C=1$. In other words, the {\em electronic content} of both $\Lc$ and $\Lc^\dagger$ is exactly the same,
except for the fact that, in this second circuit, $L$ and $C$ are fixed, while $R$ is not. Moreover, it is not difficult to extend these
results in order to get rid of the constraint $L=C=1$. The only difference is that we should identify $x_3(\tau)$ not with $-V_1(\tau)$, but
with $-LV_1(\tau)$ and $x_4(\tau)$ with $-LV_2(\tau)$. We can understand this sort of electronic equivalence between $\Lc$ and $\Lc^\dagger$
simply recalling that there exists a similarity transformation, implemented by the self-adjoint operator $S_\varphi$, which maps $\Lc$ into $\Lc^\dagger$ and viceversa.

\vspace{2mm}

\begin{figure}[h]
\begin{center}
\includegraphics[height=5.2cm, width=9.5cm]{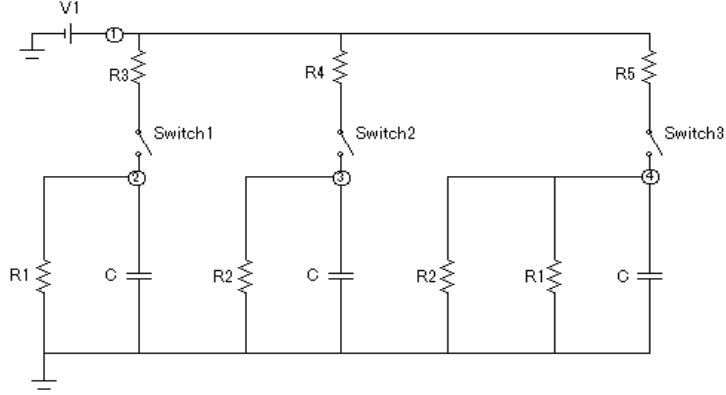}\hfill\\
\caption{\label{circuito2}\footnotesize The circuit for $H_0$.}
\end{center}
\end{figure}

{\bf Remark:--} If we repeat a similar treatment for $H_0$, which is again related to $\Lc$ and $\Lc^\dagger$, as in (\ref{extra1}), we get the differential equation $
Y'(\tau)=H_0\, Y(\tau)$, with  $Y^T(\tau)=(y_1(\tau),y_2(\tau),y_3(\tau),y_4(\tau))$, and the solution is trivial: \be y_1(\tau)=y_1(0),\quad y_2(\tau)=e^{\lambda_1\tau}y_2(0),
\quad y_3(\tau)=e^{\lambda_2\tau}y_3(0),\quad y_4(\tau)=e^{(\lambda_1+\lambda_2)\tau}y_4(0). \label{312}\en This suggests that an equivalence
between the original gain-loss circuit and a simple circuit implementing (\ref{312}), see Figure \ref{circuito2} could be established, not only
at a mathematical, but also at an electronic level. The difference between the circuits for $\Lc$, $\Lc^\dagger$ and $H_0$ could be related to
the fact that, as we have seen, the intertwining operator between $\Lc$ and $\Lc^\dagger$ is self-adjoint, while the one between $\Lc$ and $H_0$ is not.

\subsection{Heisenberg-like dynamics}

In \cite{bagpf2} we have briefly discussed that, when dealing with the time evolution of a quantum system driven by a non self-adjoint
hamiltonian, the natural choice of the   Heisenberg dynamics is not the standard $X(t)=e^{iHt}X(0)e^{-iHt}$, since this choice does not
preserve the independence of the mean values of the observables with respect to the representation chosen. The choice we made, which also
agrees with the choice made by other authors, see for instance \cite{graefe} and references therein, is the following: since the wave function
of a system, $\Phi(t)$, satisfies the equation $i\dot \Phi(t)=H\Phi(t)$, where $H$ could be self-adjoint or not, we put $$ X(t)=e^{iH^\dagger
t}X(0)e^{-iHt}, $$ for each observable $X$ of the system. In this way we have that
$\langle\Phi(t),X(0)\Phi(t)\rangle=\langle\Phi(0),X(t)\Phi(0)\rangle$. We adopt here this same recipe, identifying $H$ with $i\Lc$, as suggested in Section II. Then, after
few computations, we deduce that \be X(\tau)=e^{2l_3\tau}e^{\tilde\Lc^\dagger\tau}X(0)e^{\tilde\Lc\tau}, \label{36}\en for each operator $X$ of
the circuit. In particular, if we look for the time evolution of the number operators $N_1$ and $N_2$, using the expansion $e^{\hat \alpha
N_j}=\1+(e^{\hat\alpha}-1)N_j$, $j=1,2$ and $\hat\alpha\in\Bbb R$, and its adjoint, we find: $$
N_1(\tau)=e^{(2l_3+\lambda_1)\tau}\left(\1+(e^{\lambda_1\tau}-1)N_1^\dagger\right)N_1\left(\1+(e^{\lambda_2\tau}-1)(N_2+N_2^\dagger)+(e^{\lambda_2\tau}-1)^2N_2^\dagger
N_2\right), $$ and $$
N_2(\tau)=e^{(2l_3+\lambda_2)\tau}\left(\1+(e^{\lambda_2\tau}-1)N_2^\dagger\right)N_2\left(\1+(e^{\lambda_1\tau}-1)(N_1+N_1^\dagger)+(e^{\lambda_1\tau}-1)^2N_1^\dagger
N_1\right). $$ Since $\|N_j\|=1$, and $\lambda_j>0$, $j=1,2$, we can check that $\|N_j(\tau)\|\leq e^{-2l_3\tau}$, $j=1,2$. Recalling now that $l_3<0$, this
inequality can be used to give an upper bound on the possible growth of the operators $N_1(\tau)$ and $N_2(\tau)$. It could be worth noticing that $N_j(\tau)$ is \underline{not} explicitly related to the $j-$th sub-circuit, so that we cannot use the above formulas to deduce the time evolutions of the two gain-loss parts of the original circuit.

\section{Conclusions}

We have shown how a general framework, originally proposed in a quantum mechanical settings, can be used in the analysis of an electronic
circuit. In particular we have shown that the dynamical behavior of a gain-loss circuit can be analyzed by means of two-dimensional
pseudo-fermionic operators. In our opinion, this approach is interesting at least for two reasons:
\begin{itemize}

\item first for a purely mathematical reason: out of our simple circuit, we have produced two sets of biorthogonal bases of $\Hil={\Bbb C}^4$
having a lot of nice properties. For instance, they are related by an intertwining operator, which is the same operator which can be used
to make the Liouvillan of the circuit self adjoint;

\item from an applicative point of view, we have seen how pseudo-fermions can be useful to solve the differential equations for the circuit,
and we have also shown that other circuits can be constructed starting from the original one.

\end{itemize}

In our opinion, these results open new interesting research lines. In particular, a natural question is about some general relation, if any, between
other kinds of circuits and pseudo-fermion operators. Or, stated in different terms: for what kind of circuits a pseudo-fermionic structure can
be found? And, viceversa, given some pseudo-fermion operators and some non self-adjoint hamiltonian constructed out of them, is there any
electronic circuit which {\em implements} the dynamics? A deeper understanding of the relations, if any, between the two circuits in Figures
\ref{circuito} and \ref{circuito2} is also worth. Needless to say, a comparison between ours and the results in \cite{circu1,circu2,circu3} is also worth. These, we believe, are interesting open questions which will be considered in a near future.

\section*{Acknowledgements}

This work was partially supported by the University of Palermo.


\vspace{8mm}

 \appendix

\renewcommand{\theequation}{\Alph{section}.\arabic{equation}}

 \section{\hspace{-.7cm}ppendix: a different look to (\ref{22})}

Rather than recasting equation (\ref{22}) as in (\ref{24}), we can deduce, out of that system, two uncoupled fourth-order differential
equations for $V_1(\tau)$ and $V_2(\tau)$. In fact, it is possible to check that they both satisfy the same equation \be
v^{(iv)}(\tau)+v^{(ii)}(\tau)(2\alpha-\gamma^2)+v(\tau)\alpha^2(1-\mu^2)=0. \label{a1}\en Of course, in order to get a single solution of this
equation, we have to deduce the initial conditions for $V_j(0)$ and its first three derivatives, $j=1,2$. These will be different for
$V_1(\tau)$ and $V_2(\tau)$, so that different behavior will be deduced for the two functions even if they satisfy the same equation.

Rather than deriving the solution of this equation, we just stress here that the wave-function in (\ref{35}) can be checked explicitly to be a solution of (\ref{a1}), as it
should be, due to the uniqueness of the solution.

\end{document}